\documentclass[pss]{wiley2sp}
\usepackage{amsmath}
\usepackage{bm}              
\usepackage{w-greek}         %

\begin{document}

\title{Magnetic impurity transition in a d+s~-~wave superconductor}

\titlerunning{Magnetic impurity transition in a $d+s$-wave superconductor}

\author{%
  L.~S.~Borkowski\textsuperscript{\Ast}
}
\authorrunning{L.~S.~Borkowski}

\mail{e-mail
  \textsf{lech.s.borkowski@gmail.com}
}

\institute{%
Quantum Physics Division, Faculty of Physics, A. Mickiewicz University, Umultowska 85, 61-614 Poznan, Poland
}


\pacs{74.20.Rp,74.25.Dw,74.62.Dh,74.72.-h} 

\abstract{%
%
%
%
\abstcol{
We consider the superconducting state
of $d+s$ symmetry with finite concentration of Anderson
impurities in the limit $\Delta_s/\Delta_d \ll 1$.
The model consists of a BCS-like term in the Hamiltonian
and the Anderson impurity treated in the self-consistent
large-$N$ mean field approximation.
}{
Increasing impurity concentration or lowering the ratio $\Delta_s/\Delta_d$
drives the system through a transition from a state with two sharp
peaks at low energies and exponentially
small density of states at the Fermi level to one with
$N(0) \simeq (\Delta_s/\Delta_d)^2$. This transition is discontinuous
if the energy of the impurity resonance is the smallest
energy scale in the problem.
}
}

%
%

\maketitle   

\section{Introduction}

The order parameter symmetry of high-$T_c$ compounds
is believed to be predominantly of $d$-wave type.
However a subdominant $s$-wave component of the order
parameter may also occur, especially in materials
with orthorombic distortions.
One of such compounds is YBCO, exhibiting
strong structural distortion and a substantial
anisotropy in the London penetration depth in the $a$-$b$
plane.\cite{Basov95}
Raman scattering\cite{Limonov98} provided evidence
for a $5\%$ admixture of $s$-wave component
while thermal conductivity measurements in rotating
magnetic field\cite{Aubin97} placed an upper limit
of $10\%$.
Angle-resolved photoemission spectroscopy\cite{Lu2001}
on monocrystalline YBCO gave the ratio of 1.5 for gap amplitudes
in the $a$ and $b$ directions in the $CuO_2$.
Measurements of a Josephson current between monocrystalline
\textrm{YBa$_2$Cu$_3$O$_7$} and $s$-wave Nb showed that
the obtained anisotropy could be explained
by a $83\%$ $d$-wave with a $17\%$ $s$-wave
component.\cite{Smilde2005}
In another experiment on YBCO/Nb junction rings the $s$ to $d$ gap
ratio in optimally doped YBCO was estimated to be 0.1.\cite{Kirtley2006}

Inelastic neutron scattering on monocrystalline and untwinned
samples of YBCO lead to magnetic susceptibilities
with intensities and line shapes
breaking the tetragonal symmetry.\cite{Mook2000,Stock2004,Stock2005,Hinkov2004}
It was shown that these data may be interpreted
within an anisotropic band model with an order parameter of mixed
$d$ and $s$ symmetry.\cite{Sigrist2006}

Superconducting states with mixed symmetry are also considered
in other classes of compounds, e.g.
in the recently discovered ferropnictides.\cite{Hirschfeld2009}

The effects of dilute concentrations
of magnetic and nonmagnetic point defects
on a BCS superconductor of pure $s$ or $d$-wave symmetry
were intensively studied in the past and are well known.
In a $d$-wave superconductor with lines of order parameter
nodes any amount of disorder induces a nonzero density
of states at the Fermi level. In an $s$-wave system only
magnetic impurities change the response of the superconducting
state. For sufficiently strong coupling between
the impurity and the conduction band bound states may appear
in the energy gap.\cite{Borkowski1994}

In an earlier work on nonmagnetic
impurities in a {$d+s$}-wave superconductor it was shown
that in the unitary limit
a nonzero density of states (DOS) at the Fermi level
appears above certain critical impurity concentration,
depending on the size of an $s$-wave component.\cite{kim}
However the low-energy DOS is mostly featureless since
the $s$-wave component prevents a buildup of states
due to nonmagnetic scattering.
In contrast the presence of magnetic impurities
in a superconductor with an $s$-wave component
may result in sharp peaks in the low-energy DOS
for small concentration of defects, provided
the energy scale associated with the impurity resonance
is small.

\section{Model and Results}

We consider the order parameter of $d+s$ symmetry on a cylindrical
Fermi surface.

\begin{equation}
\Delta_s+e^{i\theta}\Delta_d(\hat k) ,
\end{equation}
where $\Delta_s$ and $\Delta_d$ are amplitudes of $s$- and $d$-wave component
respectively. We assume $\theta=0$ and $\Delta_s \ll \Delta_d$.

The superconductor is treated in a BCS approximation.
The magnetic scatterer is modelled as
an Anderson impurity treated within the slave boson mean field
approach.\cite{Borkowski2008}
The low energy physics is dominated by the presence
of strongly scattering impurity resonance. The self-consistent
self-energy equations describing the interplay between the superconducting
and magnetic degrees of freedom have the following form,

\begin{equation}
\label{omtil}
\widetilde{\omega} = \omega + {\frac{nN}{2\pi N_0}}
\Gamma \frac{\bar{\omega}}{(-\bar{\omega}^2+\epsilon^2_f)} \quad ,
\end{equation}

\begin{equation}
\label{dtil}
\widetilde{\Delta} = \Delta_s + {\frac{nN}{2\pi N_0}}
\Gamma \frac{\bar{\Delta}}{(-\bar{\omega}^2+\epsilon^2_f)} \quad ,
\end{equation}

\begin{equation}
\bar{\omega} = \omega + \Gamma
\langle \frac{\widetilde{\omega}}{\left(\widetilde{\Delta}^2(k)
- {\widetilde{\omega}}^2\right)^{1/2}}\rangle \quad ,
\end{equation}

\begin{equation}
\bar{\Delta}=\Gamma
\langle\frac{\widetilde{\Delta}}
{\left(\widetilde{\Delta}^2(k)
- {\widetilde{\omega}}^2\right)^{1/2}}
\rangle \quad ,
\end{equation}
where $\widetilde{\omega} (\bar{\omega}), \widetilde{\Delta}
(\bar{\Delta})$ is the renormalized frequency and order parameter
of conduction electrons (impurity) respectively.

We assume $\Delta_d/D=0.01$,
where $2D$ is the bandwidth of the conduction electron band.
In the equations above $\Gamma$ is the hybridization energy
between the impurity and the conduction band and
$\epsilon_f$ is the resonant level energy. Here we assume
constant density of states in the normal state
$N_0=1/2D$ and do the calculations for a nondegenerate
impurity, $N=2$.
Brackets denote average over the Fermi surface.

Initial results for the density of states of this system
were presented in an earlier paper.\cite{Borkowski2008a}
For small impurity concentration $n$, such that $\Delta_s$
and $\Delta_d$ are not significantly affected,
there are two peaks located symmetrically
near the gap center, provided $\epsilon_f \ll \Gamma \ll \Delta_s$.
For larger $n$ the two peaks
merge into one peak centrally located at the Fermi energy.

\small
\begin{figure}[th]
\centering
\includegraphics[width=20pc]{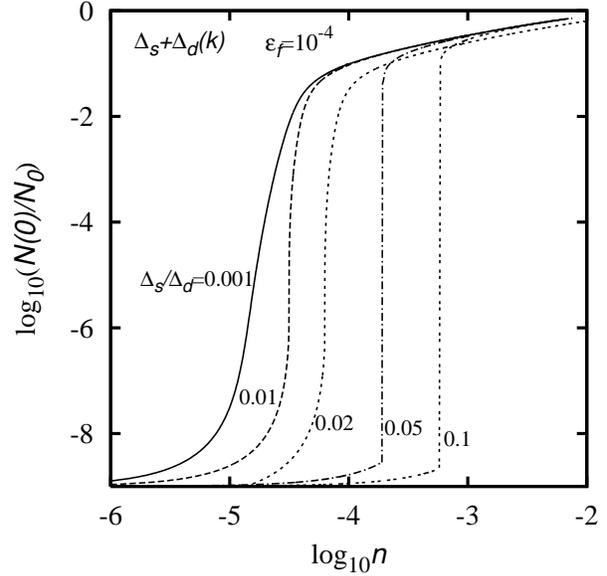}
\caption{Logarithm of the density of states at the Fermi level
as a function of impurity concentration for several values
of the ratio $\Delta_s/\Delta_{d0}$. The resonant impurity level
$\epsilon_f$ is close to the Fermi level. $\Gamma/D$ is fixed
at 0.001.
}
\label{fig1}
\end{figure}
\normalsize

\small
\begin{figure}[th]
\centering
\includegraphics[width=20pc]{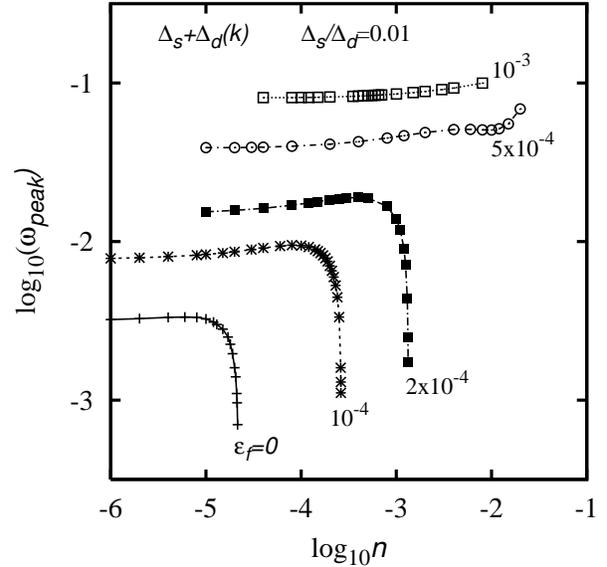}
\caption{
Position of the peak in the conduction electron DOS as a function
of impurity concentration for $d+s$ superconductor for several
values of the resonant level energy. Energy is scaled by half
of the conduction electron bandwidth $D$ and $\Gamma/D=0.001$.
The suppression of $\Delta_s$ and $\Delta_d$
was not taken into account.}
\label{fig2}
\end{figure}
\normalsize

Fig. 1 shows the dependence of the density of states $N(0)$
at $E_F$ as a function of impurity concentration.
At small $n$, $N(0)$ is exponentially small,
$N(0)/N_0 \sim (\Delta_s/\Delta_d)
\exp(-\alpha (\Delta_s/\Delta_d)^2/n)$, where
$\alpha$ is a numerical factor.
The critical concentration $n_0$ for the discontinuous
transition to $N(0)/N_0 \simeq \Delta_s/\Delta_d$
is a quadratic function of $\Delta_s/\Delta_d$.
For $n > n_0$, $N(0)$ approaches DOS
of a $d$-wave superconductor in both the magnitude and its
functional dependence on $n$.
These relations are valid when $\epsilon_f \ll \Gamma, \Delta_s$.

Qualitatively similar scaling of $N(0)$ as a function
of $\Delta_s/\Delta_d$ in presence
of nonmagnetic impurities in the unitary
limit was obtained in ref. \cite{kim}.
The difference between magnetic and nonmagnetic
defetcs in a superconductor with $\Delta_s/\Delta_d \ll 1$
is the character of the transition from the exponentially small $N(0)$
to finite $N(0)$ in the strong scattering limit.
In contrast to nonmagnetic impurities, the transition caused
by resonantly scattering magnetic impurities is discontinuous.
and there are two sharp peaks
of $N(\omega)$ on both sides of the transition.\cite{Borkowski2008a}
The position of peaks as a function of impurity concentration
for different values of $\epsilon_f$ is shown in Fig 2.
These peaks strongly alter the low-energy and low-frequency
response and may be detected in thermodynamic or transport measurements.
The increase of $\Delta_s$ shifts the resonances towards $E_F$
and makes them more narrow.

The $d$-wave component of the superconducting order
parameter is more sensitive to pair breaking
by magnetic impurities in the limit $T \ll T_K$,
where $T_K=\sqrt{\Gamma^2+\epsilon_f^2}$,
than the $s$-wave part.
Depending on the relative size of $\Delta_d$ and $\Delta_s$
there may be another impurity
transition for larger $n$, when the order
parameter nodes disappear due to vanishing of the $d$-wave
component and a full
gap opens up. The impurity peak is then split
and $N(0)$ falls to zero again.
Hower a detailed description of this possibility requires
a careful analysis involving four energy scales:
$\Delta_d$, $\Delta_s$, $\Gamma$, and $\epsilon_f$.

Can such transition be observed experimentally?
The transition may be tuned either by varying impurity concentration
or the ratio $\Delta_s/\Delta_d$.
The density of states of the normal state
at $E_F$, $N_0$, also has an effect.
It appears in equations~\ref{omtil} and \ref{dtil}.
One should bear in mind, however, that
both the superconducting transition temperature
and the impurity resonance energy scale $T_K$
are exponentially sensitive to changes of $N_0$.
Experiments conducted at very low temperatures
may give different signatures of low-energy behavior
depending on the location of the system on the phase diagram
relative to the impurity critical point.

In a $d+s$-superconductor in the limit
of high concentration
of point defects the $d$-wave component vanishes
while the $s$-wave part of the order parameter remains
unaffected.
The situation is different in presence of magnetic
scatterers. If the $s$-wave component is small
and the energy scale of the resonance due to impurity
scattering is at most of the order $\Delta_s$,
increasing impurity concentration
may drive $\Delta_s \rightarrow 0$ while
$\Delta_d$ will be reduced and finite.
This follows from the fact that the largest pair breaking
occurs when the energy scales of the impurity resonance
and the order parameter are comparable.
If $\Delta_d \gg T_K$,
the rate of suppression
of $\Delta_d$ with increasing $n$ is small.

The $d+s$ superconductor with a subdominant $s$-wave
component doped with magnetic impurities
has a different phase diagram in the $T$-$n$ plane
compared to the same superconductor with nonmagnetic defects.
If the impurity resonance scale $T_K$ is comparable to $\Delta_s$
and $\Delta_s \ll \Delta_d$, the $s$-wave component
may vanish at $T \simeq T_K$.

\section{Conclusions}
The $T=0$ impurity transition
discussed in this work
may be detected in the low temperature limit.
While the slave boson mean field formalism used in this paper
cannot be applied at temperatures exceeding $T_K$,
we may also qualitatively describe the expected behavior of the system
as a function of temperature.
At finite but small impurity concentration
there may be even four phase transitions as a function
of temperature: normal to $d$-wave, $d$-wave to $d+s$-wave,
$d+s$ to $d$-wave, and $d$-wave to $d+s$-wave.
Due to the difference of magnitudes
the small $\Delta_s$ may
be driven to zero faster with increasing $n$ than $\Delta_d$.
Nonmonotonic behavior around $T \sim T_K$
is a consequence of strong scattering by the resonance
state forming on the impurity site.
As $T \rightarrow 0$, the impurity scattering
becomes weaker and the $s$-wave component may appear again.
Whether this particular scenario is realized, depends
on the relative size of energy scales:
$\Delta_s/\Delta_d$, $\Delta_s/T_K$,
and $\epsilon_f/\Gamma$.

\begin{acknowledgement}
Some of the computations were performed in the Computer Center
of the Tri-city Academic Computer Network in Gdansk.
\end{acknowledgement}

%
%

\end{document}